\documentstyle[prl,aps,preprint,epsfig]{revtex}
\begin{document}
\tighten

\title{Drell-Hearn-Gerasimov Sum Rule for \\
       the Nucleon in the Large-$N_c$ Limit}
\author{Thomas D. Cohen and Xiangdong Ji}
\bigskip

\address{
Department of Physics \\
University of Maryland \\
College Park, Maryland 20742 \\
{~}}

\date{UMD PP\#99-108 ~~~DOE/ER/40762-181~~~ May 1999}

\maketitle

\begin{abstract}
We show that the Drell-Hearn-Gerasimov sum rule
for the nucleon is entirely 
saturated by the $\Delta$ resonance
in the limit of a large number of colors, $N_c
\rightarrow \infty$. Corrections are at
relative order $1/N_c^2$. 

\end{abstract}
\pacs{xxxxxx}

\narrowtext
The Drell-Hearn-Gerasimov (DHG) sum rule for a spin-1/2 
fermion relates its anomolous magnetic moment to an 
integral over the spin-dependent photo-production cross 
section\cite{DHG}. Define the total photo-production 
cross sections $\sigma_P(\nu)$ and $\sigma_A(\nu)$, where 
$P$ and $A$ label the spins of the photon being 
parallel and antiparallel, respectively, to the spin of 
the fermion and $\nu$ is the photon energy. The 
DHG sum rule says,
\begin{equation}
    \int^\infty_{\nu_{\rm in}}
      d\nu{\sigma_P(\nu)-\sigma_A(\nu)\over \nu}
    = {2\pi^2\alpha_{\rm em} \kappa^2\over M^2} \ . 
\label{DHG}
\end{equation}
where $\kappa$ and $M$ are the anomalous magnetic moment
(dimensionless) and mass of the fermion, respectively.  
For the case of the nucleon it is useful to break the 
sum rule up into isoscalar and isovector components.

It has been known empirically for a long time
that the isoscalar part of the DHG sum rule for
the nucleon is almost entirely dominated by 
the $\Delta$-resonance contribution. 
The theoretical expectation for the sum ({\it i.e.} 
the right-hand side (RHS) of eq.~(\ref{DHG})) 
is 219 $\mu b$, whereas 
the $\Delta$-resonance contribution to the left-hand side (LHS)
integral is 240$\mu b$ \cite{karliner}.
To the best of our knowledge, there hasn't been 
a solid theoretical explanation for this from fundamental principles
of quantum chromodynamics (QCD). In this short note,
we show that such phenomenon is expected in QCD 
in the limit of a large number of colors ($N_c\rightarrow\infty)$
\cite{thooft,Witten}. 

During the past several years, there has been significant progress
in understanding the baryon properties in the 
large-$N_c$ QCD. Large-$N_c$ consistency of the theory
has been exploited to show that it 
has a contracted $SU(2n_f)$ spin-flavor symmetry, 
where $n_f$ is the number of quark flavors 
\cite{GS,manohar}. For example, the symmetry 
is required for cancellations among Feynman 
diagrams with different intermediate states in a 
hadronic description of scattering in order to preserve 
the unitarity bound. The spin-flavor symmetry 
implies that there is a tower 
of states with $I=J$ which are degenerate in the 
large-$N_c$ limit; the lowest two states
are identified with the nucleon and the $\Delta$-resonance. 
It has also been shown the splitting between the lowest states 
in this tower ({\it eg.} the nucleon and $\Delta$) scales as 
$1/N_c$ \cite{Jenkins}. The symmetry also 
has the feature that certain hadronic matrix elements are 
related to each other in the large-$N_c$ limit.  
For example, the ratio of the axial-current matrix 
element in the nucleon to the transition matrix element 
between the nucleon and the $\Delta$ is fixed in the large-$N_c$ 
limit. Moreover, there are cases, such as the
axial matrix elements, where the leading correction to this 
ratio occurs at relative order $1/N_c^2$ \cite{manohar}. 

As pointed out by Broniowski \cite{bro}, the same underlying 
symmetry principle is responsible for understanding large-$N_c$ 
consistency of low-energy sum rules such as those of 
Addler-Weisberger\cite{AW} and Cabibbo-Radicati \cite{CR}.  
In this paper, we follow Broniowski's observation 
and point out that the $\Delta$-dominance in the 
isoscalar part of the DHG sum rule is required by 
a similar large-$N_c$ consistency and reflects 
the above-mentioned spin-flavor symmetry.  Moreover, we will 
show that
the $\Delta$ dominance is valid up to relative order $1/N_c^2$.

Let us do some large-$N_c$ counting for the electromagnetic
couplings in order to understand the large-$N_c$ consistency 
of the DHG sum rule. Consider two quark flavors:
up and down. The proton (neutron) 
is then made of $(N_c\pm 1)/2$ up and $(N_c\mp 1)/2$
down quarks. We adopt a scheme of assigning electric
charges to the quarks such that the proton and neutron
have the usual charges one and zero. Thus, 
$e_{u(d)} = \pm 1/2+1/2N_c$. Other charge-assignment 
schemes are possible, such as keeping the quark charges fixed 
at 2/3 and $-1/3$. We will not explore them
here because they do not affect the main conclusion 
of the paper. In the large $N_c$-limit,
the quark matrix elements in the nucleon have the
following simple $N_c$-counting rules \cite{manohar}:
\begin{eqnarray}
        \left\langle \sum_i 1\right\rangle &\sim& N_c\ ,   \nonumber \\
        \left\langle \sum_i \vec{\tau_i} \right\rangle &\sim& 1 \ , 
         \nonumber \\
        \left\langle \sum_i \vec{\sigma_i} 
           \right\rangle &\sim& 1 \ ,  \nonumber \\
        \left\langle \sum_i \vec{\tau_i}\vec{\sigma_i} \right\rangle &\sim& N_c \ . 
\end{eqnarray}
where $\vec{\sigma_i}$ and $\vec{\tau_i}$ are the
spin and isospin matrices of the $i$-th quark. 
According to the above, the $N_c$-counting for the 
electromagnetic couplings of the nucleon goes as
\begin{eqnarray}
        {\rm isovector~ charge~ coupling}~~ &e_V~ &\sim ~1 \ ,  \nonumber \\
        {\rm isoscalar~ charge~ coupling}~~ &e_S~ &\sim ~1 \ ,  \nonumber \\
        {\rm isovector~ magnetic~ coupling}~~ &\mu_V& ~\sim~ N_c \ , \nonumber \\
        {\rm isoscalar~ magnetic~ coupling}~~ &\mu_S& ~\sim~1/N_c \ . 
\end{eqnarray}
The preceding  counting rules appear to be results of the simple quark model.
In fact, they are general results of the group theory---the quark model 
language is serving as a simple device to do the group theory
conveniently.

The isovector magnetic coupling scales as $N_c$.
This leads to a rather perverse formal large-$N_c$
limit.  As $N_c \rightarrow \infty$, the electromagentic
energy of the nucleon goes as $\alpha_{\rm em} N_c^2$. Keeping
$\alpha_{\rm em}$ at order 1, we see that the electromagnetic
energy will eventually become nonperturbative and overwhelm 
the strong interaction contribution which is order $N_c$. 
In practice we wish to study a regime close to the real 
world in which electromagnetism is, indeed, perturbative. 
A simple way to do this is to take  
$\alpha_{\rm em }N_c$ to be independent of $N_c$ and $<\!\!<1$,  
which is well satisfied in the real world.  

Now we are ready to examine the large-$N_c$
consistency of the DHG sum rule. First, consider
the RHS of Eq. (1) 
in the large-$N_c$ limit. According to the previous 
discussion, its leading dependence in our expansion is 
$\alpha_{\rm em} N_c^2 $ because 
$\kappa_V/M \sim \mu_V \sim N_c$. 

Second, it is easy to see that the integrand of
the spin-dependent cross section in the 
LHS of the DHG sum rule (Eq. (1)) generically 
as $\alpha_{\rm em} N_c^0$. Indeed, in the quark representation,
\begin{equation}
   \sigma_{P}-\sigma_{A} \sim \alpha_{\rm em} \left\langle
      \sum_i \vec{\sigma_i} \right\rangle\ , 
\end{equation}
which according to Eq. (2), scales like $\alpha_{\rm em} N_c^0$. 

So, how does the DHG sum rule work 
if the two sides  have different large-$N_c$  behavior?
The answer lies in the $\Delta$-resonance 
contribution.  While the generic contribution to the LHS is 
${\cal O}( \alpha_{\rm em} N_c^0)$, the $\Delta$ 
contribution, as a member of the same tower of states 
as the nucleon, is special. In the large-$N_c$ limit, 
the $\Delta$-resonance is the only state excitable
with a photon of energy of order $1/N_c$. Moreover,
the $\Delta$-nucleon transition magnetic moment, like the
nucleon's isovector magnetic moment is order $N_c$.  
Overall, the $\Delta$ contribution to 
the spin-dependent cross section turns out to be quite special:
it is of order $\alpha_{\rm em} N_c$. Therefore, 
the $\Delta$-resonance contribution to the DHG sum rule 
can be of order $\alpha_{\rm em} N_c^2$.  

Indeed, a straightforward calculation yields the
following $\Delta$-photoproduction cross section, 
\begin{equation} 
   \sigma_\Delta(\nu) = \delta(\nu-\Delta) \Delta {2\pi^2\alpha_{\rm em}
\mu^2_{\gamma\Delta N}\over M^2} \ ,
\end{equation}
where $\Delta=M_\Delta-M$ is the nucleon-$\Delta$ mass difference
and $\mu_{\gamma\Delta N}$ is the M1 coupling between
the photon, nucleon and $\Delta$-resonance. 
Its contribution to the LHS of the DHG integral is
\begin{equation}
     {2\pi^2\alpha_{\rm em} \mu^2_{\gamma\Delta N}
    \over M^2} \ .    
\end{equation}
In the large $N_c$ limit $\mu_{\gamma N\Delta}
= \kappa_V$, and so the above contribution 
matches exactly the leading $N_c$ contribution
from the RHS of Eq. (1). Since the isoscalar anomalous 
magnetic moment $\kappa_S/M$ is of order $1/N_c$, 
the relation $\mu_{\gamma\Delta N}
=\kappa_V$ is in fact correct up to ${\cal O}(N_c)$
($\kappa_V$ is ${\cal O}(N_c^2)$). 

The large-$N_c$ consistency of the
isovector component of the DHG sum rule
is simple: The RHS of the sum rule
scales like $\alpha_{\rm em}\kappa_V\kappa_S/M^2
\sim \alpha_{\rm em}$. The isovector spin-dependent
cross section goes like $\alpha_{\rm em}\langle
\sum_i \vec{\sigma_i}\vec{\tau_i}\rangle/N_c \sim 
\alpha_{\rm em}$; the $N_c$ counting is completely consistent.
 
The Drell-Hearn-Gerasimov sum rule is derived 
under the condition that $N_c\ne \infty$. However, 
if one starts with QCD with $N_c=\infty$, then the low-energy
theorem for the Compton scattering amplitude 
changes since the $\Delta$ is now a massless excitation.
The Compton amplitude, $S_1(\nu, Q^2)$, is defined according to 
\begin{eqnarray}
 T^{\mu\nu} &= &i \int d^4\xi e^{i\xi \cdot q} \langle
PS|TJ^\mu(\xi)J^\nu(0)|PS\rangle \nonumber \\
 & = & -i\epsilon^{\mu\nu\alpha
\beta}q_\alpha S_\beta S_1(\nu, Q^2)+... \ , 
\end{eqnarray}
where $S^\alpha$ is the polarization vector of the
nucleon. Including the $\Delta$-resonance contribution, 
we find
\begin{equation}
     S_1(0,0) = -({2\kappa_S\kappa_V+\kappa_S^2})/M^2 \ . 
\end{equation}
The leading-order nucleon-pole contribution is exactly cancelled 
by the $\Delta$-resonance pole. This cancellation reflects
the contracted $SU(4)$ symmetry.
On the LHS of the DHG, the $\Delta$-resonance
contribution is absent because it is forbidden by energy-momentum
conservation. 

It is significant to notice that the 
$\Delta$ dominates the DHG sum rule
up to corrections of order $\alpha_{\rm em} 1/N_c^2$.  
The easiest way to 
see this is to compare the two sides of the sum rule.  The
right-hand side scales as $\alpha_{\rm em} N_c^2$ 
while the generic contributions to the left-hand side are order 
$\alpha_{\rm em} N_c^0$.  One
concludes that the $\Delta$ contribution must therefore account for 
both the $\alpha_{\rm em} N_c^2$ and $\alpha_{\rm em} 
N_c$ contribution on
the left-hand side.  In fact, the lack of a correction 
at relative order $1/N_c$ is easily understood in another way.  
As noted above, there are cases, such as the axial-vector current, 
where the ratios of nucleon matrix elements and 
nucleon-$\Delta$ transition matrix elements 
are fixed with corrections at 
relative order $1/N_c^2$\cite{manohar}.  The isovector magentic
 coupling of a photon has the same structure (in terms of the 
couplings
 to quarks) as the axial current.  Thus the ratio of the 
nucleon isovector
 magnetic moment and the 
nucleon-$\Delta$ magnetic transition moment is 
also fixed by the group structure
up to relative corrections of order $1/N_c^2$. 

To summarize, we argued that 
the DHG sum rule is saturated by the 
$\Delta$ resonance in the large-$N_c$ limit up to correction terms of
relative order $1/N_c^2$.
The result reflects the contracted spin-flavor $SU(4)$
symmetry in the large-$N_c$ QCD with two 
flavors. Turning the argument around, the contracted
$SU(4)$ symmetry indicates that the spin-dependent
photonucleon cross section must be of order
$\alpha_{\rm em}N_c^0$. 

\vskip 0.5in

We thank C.-W. Kao and J. Osborne for discussions
about the subject. This work is supported 
by funds provided by the U.S.  Department of Energy 
(D.O.E.) under cooperative agreement DOE-FG02-93ER-40762.


\begin{references}
\frenchspacing

\bibitem{DHG}
S. D. Drell and A. C. Hearn, Phys. Rev. Lett. {\bf 16} (1966) 908;\\
S. B. Gerasimov, Sov. J. Nucl. Phys. {\bf 2} (1966) 430. 

\bibitem{karliner}
I. Karliner, Phys. Rev. {\bf D7} (1973) 2717; 
R. L. Workman and R. A. Arndt, Phys. Rev. {\bf D45} (1992) 1789. 

\bibitem{thooft}
G. t' Hooft, Nucl. Phys. {\bf B72} (1974) 461. 

\bibitem{Witten}
E. Witten, Nucl. Phys. {\bf B160} (1979) 57; {\bf B156} (1979) 269.

\bibitem{GS}
J. Gervais and B. Sakita, Phys. Rev. Lett. {\bf 52} (1984) 87. 

\bibitem{manohar}
R. Dashen, E. Jenkins, and A. Manohar, Phys. Rev. {\bf D49} 
(1994) 4713; {\bf D51} (1995) 3697. 

\bibitem{Jenkins}
E. Jenkins, Phys. Lett. {\bf B315} (1993) 441; 
Phys. Rev. {\bf D54} (1996) 4515; \\
E. Jenkins and R. F. Lebed, Phys. Rev. {\bf D52} (1995) 282. 

\bibitem{bro}
W. Broniowski, Nucl. Phys. {\bf A580} (1994) 429. 

\bibitem{AW}
A. Adler, Phys. Rev. {\bf 140} (1965) 736; \\
W. I. Weisberger, Phys. {\bf 143} (1966) 1302. 

\bibitem{CR}
N. Cabibbo and L. A. Radicati, Phys. Lett. {\bf 19} (1966) 697. 

\bibitem{ji}
X. Ji and J. Osborne, to be published. 

\nonfrenchspacing
\end{references}
\end{document}